# On the role of Initial Error Growth in the Skill of Extended Range Prediction of Madden-Julian Oscillation (MJO)

Lekshmi S[1,2], Rajib Chattopadhyay[1], Manpreet Kaur[1], Susmitha Joseph[1], R.Phani[1], A Dey[1], R. Mandal[1], AK. Sahai[1]

1. Indian Institute of Tropical Meteorology, Pune-411008
2. Cochin University of Science and Technology

# **ABSTRACT**

The seamless forecast approach of subseasonal to seasonal scale variability has been succeeding in the forecast of multiple meteorological scales in a uniform framework. In this paradigm, it is hypothesized that reduction in initial error in dynamical forecast would help to reduce forecast error in extended lead-time up to 2-3 weeks. This is tested in a version of operational extended range forecasts based on Climate Forecast System version 2 (CFSv2) developed at Indian Institute of Tropical Meteorology (IITM), Pune. Forecast skills are assessed to understand the role of initial errors on the prediction skill for MJO. A set of lowest and highest initial day error (LIDE & HIDE) cases are defined and the error-growth for these categories are analysed for the strong MJO events during May to September (MJJAS). The MJO forecast initial errors are categorized and defined using the well-known multivariate MJO index introduced by Wheeler & Hendon (2004). The probability distribution of *bivariate RMSE* and *error growth* evolution (first order difference of index error for each successive lead days) with respect to extended range lead-time are used as metrics in this analysis. The result showed that initial error is not showing any influence in the skill of model after a lead time of 7-10 days and the error growth remains the same for both set of errors. A rapid error growth evolution of same order is seen for both the classified cases. Further the physical attribution of these errors are studied and found that the errors originate from the events with initial phase in Western Pacific and Indian Ocean. The spatial distribution of OLR and the zonal winds also confirms the same. The study emphasise the importance of better representation of MJO phases especially over Indian ocean in the model to improve the MJO prediction rather than focusing primarily on the initial conditions.

# 1. Introduction

The Madden-Julian Oscillation is a planetary scale eastward moving disturbance with an extensive zonal scale (wavenumber 1-3) and a spectral peak of 40-50 days (Madden and Julian, 1971, 1972; Zhang C, 2005), which modulates deep convection and precipitation in the tropics. It act as a dominant source of sub-seasonal predictability (Waliser, 2011) accounting for major atmospheric and oceanic variability over tropics as well as to a good extent in the extra-tropics (Zhang and Gottschalck, 2002; Fauchereau, Pohl and Lorrey, 2016). The systematic assessment of the existing methods and models in capturing different parameters associated with the propagation of MJO across the globe is vital to strengthen its prediction.

The weather-climate prediction gap at the sub-seasonal to seasonal (S2S) time scale is mainly sought by the agricultural, water management and other mitigation sectors. The steadfast research on the S2S scale made breakthrough progress in accomplishing proficient forecast (Abhilash *et al.*, 2011; Hudson *et al.*, 2011; Koster *et al.*, 2011; Wang *et al.*, 2014; H. M. Kim *et al.*, 2014; Neena *et al.*, 2014; Vitart, 2014, 2017; Robertson *et al.*, 2015; Vitart and Robertson, 2018; Dey *et al.*, 2020). Various studies conducted in the past assess the skill of different dynamical models in predicting MJO (Jones *et al.*, 2000; Vitart *et al.*, 2007; Lin, Brunet and Derome, 2008; Rashid *et al.*, 2011; Kim *et al.*, 2014). Apart from these, diverse modelling experiments assisted in assessing and comparing the forecast skill of multiple models (Kim *et al.*, 2009; Neena *et al.*, 2014; Lim, Son and Kim, 2018). Lin et al. (2008) found that the model forecasts depend on the initial phase; the forecast from the two models used in the study initialized with active convection over tropical African and Indian Ocean showed a better skill than those initialized in other phases for strong MJO events. At the same time, Rashid *et al.*, (2011) and Neena *et al.*, (2014) observed no such connection in most models. According to, Lim et al. (2018) MJO prediction skill was better for the events with a higher initial amplitude. Also, Neena *et al.*, 2014, reported the error growth on the intra-seasonal scale to be different among the different models with comparable initial error values. Therefore, the error growth at S2S scale cannot solely depend on initial conditions but could be governed by model processes.

The empirical orthogonal functions (EOF) are widely used to extract the intra-seasonal time scale variability, strongly associated with MJO. The real-time Multivariate MJO index (RMM Index) proposed by Wheeler and Hendon, 2004 (hereafter WH04), serves as a seasonally independent index for monitoring MJO. It is based on a pair of EOFs of the combined fields of near-equatorially averaged 850-hPa zonal wind (U850), 200-hPa zonal wind (U200), and the satellite observed outgoing longwave radiation (OLR). The principal component (PC) time series is extracted by projecting the daily observed data onto the EOF after removing the annual cycle and inter-annual variability. The leading two PC time series thus obtained (called the RMM indices) effectively extracts MJO signal present in the data without using a filter. Later on, several variants of these indices (Stachnik and Chrisler, 2020) are developed based on the above method for capturing the propagation of MJO. The OLR based index known as OLR-MJO index (OMI), better captures the convective signal of the MJO (Kiladis *et al.*, 2014). A velocity potential MJO Index (VPM) calculated using 200 hPa velocity potential (VP200) instead of OLR in the WH04 method aided the description of the planetary-scale characteristics of divergence associated with MJO (Ventrice *et al.*, 2013). Another operational tracking method is put forward for the smooth propagation of both MJO and the Monsoon Intra-Seasonal Oscillation (MISO) utilizing the Extended EOF analysis of U850, U200, and VP200 (Dey *et al.*, 2019).

The Climate Forecast System (CFS) is a fully coupled dynamical prediction system; its version 1 was operational at National Centre for Environmental Prediction (NCEP) from August 2004 (Saha *et al.*, 2006) replaced by the successor version (CFSv2) since March 2011

(Saha *et al.*, 2014; Jones, Hazra and Carvalho, 2015). The CFSv2 model showcase useful MJO prediction skill out to about 20 days lead; with the correlation coefficient (CC) higher than 0.8 for initial 10 days. Afterward CC drops to 0.5 and root mean square error (RMSE) increases to 1.4. (Wang *et al.*, 2014).

In the present study we will analyse the skill of the Indian Institute of Tropical Meteorology's (IITM) CFSv2 extended range prediction system (ERPS) in predicting the MJO events during the MJJAS period. The extended range prediction (ERP) group has indigenously developed ERPS and has been providing the real-time forecast of the active-break spells of Indian summer monsoon rainfall since 2011 (Abhilash *et al.*, 2011; A. K. Sahai *et al.*, 2013; Abhilash, A. K. Sahai, Borah, Chattopadhyay, *et al.*, 2014; Abhilash, Sahai, Pattnaik, *et al.*, 2014; Dey *et al.*, 2020). In the present study, we have considered the strong MJO events and has separated two set from these events; one set consists of strong events that showed the least error in PCs during the initial day forecast and the other set includes those with the highest error in initial day forecast. The error and the error growth distribution of both the set of events are investigated. Further analysis pinpoints the causes that lead to increased error in the highest initial day error cases. The objective of the study is (a) to investigate the effect of initial day error on the forecast skill of the MJO in the IITM CFSv2 model (b) to highlight the source of the large model error cropping up from the very initial lead days in the high error cases as compared to the low error cases. A similar approach, as used in WH04, is adopted here to extract the signals associated with MJO. The only exception is that instead of removing the annual cycle and inter-annual variability as in WH04, we have used a band pass filter of 30-60 days to calculate the EOF.

In Section 2, the observational and model datasets used for the study are briefly described. The methods used for obtaining the RMM indices and forecast skill are explained in Section 3. Section 4 presents the results and discussions, and the conclusion obtained from the study is given in Section 5.

## 2. Observational and Model Data

The baroclinic structure along with the convection and propagation characteristics associated with the MJO are captured using the outgoing long-wave radiation (OLR), zonal winds at 850hPa (u850) and 200 hPa (u200) levels following WH04.

### 2.1 Datasets

The daily averaged value of OLR is obtained from the National Oceanic and Atmospheric Administration (NOAA) polar orbiting satellites(Liebmann and Smith, 1996) The interpolated real-time data are acquired directly from NCEP for a period from 1979-2018. The zonal wind data used are taken from the NCEP-National Centre for Atmospheric Research (NCEP-NCAR) reanalysis datasets(Kalnay *et al.*, 1996) for the same period. All the fields are analysed on a $2.5^0 \times 2.5^0$ latitude-longitude grid.

The model data is from the CFSv2 hindcast of IITM-Extended Range Prediction System (IITM ERPS). This multi-model ensemble (MME) prediction system developed in IITM is using the NCEP's CFSv2 (Abhilash *et al.*, 2011; A. Sahai *et al.*, 2013; Abhilash, A. Sahai, Borah, Chattopadhyay, *et al.*, 2014; Abhilash, Sahai, Pattnaik, *et al.*, 2014; Saha *et al.*, 2014) and the atmospheric component, Global Forecast System (GFS); both models run at two resolutions T382 (~38 km) and T126 (~100 km) each with 4 member ensemble of perturbed initial conditions(ICs). IITM ERPAS provide real-time predictions for 32 days based on every Wednesday IC, these oceanic and atmospheric ICs for initializing the model are from INCOIS (Indian National Centre for Ocean Information Services) and NCMRWF (National Centre for Medium Range Weather Forecasting) assimilation systems respectively. The present study utilizes only 4 ensemble member of CFSv2 from IITM ERPS. The anomaly of each field is analysed in the paper by removing the corresponding model climatology using the hindcast data from 2003-2015.

## 3. Methodology

### 3.2 RMM Indices

The multivariate EOFs are computed using the OLR and wind (u850 and u200) observations from 1979-2001. The equatorially averaged ($15^0$S-$15^0$N) anomaly of these three fields is filtered using a band-pass filter of 30-60 days to isolate the signals associated with MJO. Further, each of the filtered fields is normalized by its global variance, and these combined fields are used to obtain the EOF. Following WH04, the leading EOF pair, which describes the large-scale eastward propagation of MJO, are considered, and also shown in Fig (1). The percentage of variance associated with both EOFs is shown in the top right of each EOF. EOF1 and EOF2 explain together, almost 50% of the total variance in the atmospheric field. While EOF1 manifests the enhanced convection and associated lower and upper-level wind circulation as existing over the Indian Ocean, the EOF2 is in near quadrature with EOF1.

The observational anomaly fields are projected onto the thus determined EOF pair to compute the leading pair of PCs, RMM1 and RMM2 (Real-Time Multivariate MJO Index). In Fig (2), the RMM indices are demonstrated using the phase-space diagram for the Dynamics of Madden-Julian Oscillation (DYNAMO) field campaign (October-December 2011) period (Yoneyama, Zhang and Long, 2013). In the figure, the sequential days are connected with lines that trace the anticlockwise circle indicate the eastward propagation of MJO. The model anomaly data during May to September (MJJAS) from 2003-2018 are projected onto the same EOF pair to find the model forecasted RMM indices.

## 3.3 Forecast Skill

The RMM indices can be used as a bivariate index for measuring the forecast skill of the model at any lead time τ. The bivariate coefficient of correlation (BVCC) and the bivariate root mean square error (BVRMSE), as put forward by Lin, Brunet and Derome, 2008 are used here to score the model forecasts.

$$\text{COR}(\tau) = \frac{\sum_{t=1}^{N}[a_1(t)b_1(t,\tau)+a_2(t)b_2(t,\tau)]}{\sqrt{\sum_{t=1}^{N}[a_1^2(t)+a_2^2(t)]}\sqrt{\sum_{t=1}^{N}[b_1^2(t,\tau)+b_2^2(t,\tau)]}} \quad \text{---- (1)}$$

$$\text{RMSE}(\tau) = \sqrt{\frac{1}{N}\sum_{t=1}^{N}([a_1(t)-b_1(t,\tau)]^2 + [a_2(t)-b_2(t,\tau)]^2)} \quad \text{--- (2)}$$

Here $a_1(t)$ and $a_2(t)$ are the verification RMM1 and RMM2 at any time t and $b_1(t, \tau)$ and $b_2(t, \tau)$ are the corresponding forecasts at time t for a lead time of τ days. N indicates the number of forecasts considered. The bivariate amplitude of RMM indices for verification and forecasts are given by Eq. (3) and Eq. (4)

$$\text{RMMA}_{obs}(t) = \sqrt{a_1(t)^2 + a_2(t)^2} \quad \text{--- (3)}$$

$$\text{RMMA}_{for}(t,\tau) = \sqrt{b_1(t,\tau)^2 + b_2(t,\tau)^2} \quad \text{--- (4)}$$

The error in amplitude is obtained as,

$$ERR_{amp}(\tau) = \frac{1}{N}\sum[AMP_{for}(t,\tau) - AMP_{obs}(t)] \quad \text{--- (5)}$$

## 3.4 Selection criteria of Initial conditions

This study, only considered the strong MJO events with initial PC amplitude (obtained using Eq. 3) greater than 1.0 during MJJAS. From the forecasts using 352 different initial conditions during the period mentioned above, 185 cases are identified as strong MJO events.

The error growth pattern of strong events is analysed with the help of two sets of events. These sets are obtained after arranging the events in the ascending order of error in the initial day forecast PC. Each set consist of 25% of events with Least Initial Day Error (top 46 cases) and Highest Initial Day Error (bottom 46 cases) forecast (hereafter mentioned as *LIDE* and *HIDE cases,* respectively).

## 4. Results

### 4.1 Effect of initial day error on the MJO forecast skill in the IITM CFSv2 model

The IITM CFSv2 model skill in predicting the strong MJO events during MJJAS with initial PC amplitude above 1.0 is analysed. As mentioned earlier, the group of LIDE cases and HIDE cases are compared for better insight into its role in error growth of the model forecasts. The BVRMSE and BVCC of PCs of these strong MJO events predicted by the IITM CFSv2 model are shown in Fig 2.a. The model exhibits a forecast skill of ~15 days, and after a lead time of 23 days the error does not show any marginal increase; rather remains constant. A sharp increase in error up to around 20 days is noted (Fig 2.b) from the BVRMSE for the LIDE cases; afterward, a small decrease in error is observed. The same can also be inferred from the BVCC pattern. Even though the error increases clearly with lead days, the model could capture the LIDE events during the initial few days of forecast. The rise in error is more gradual for HIDE cases (Fig 2.c) than in LIDE. However, both LIDE and HIDE error patterns seem to be more or less similar after around 15 days. During the initial lead days, the high error in PC explicates that the model cannot capture MJO even at a shorter lead for the HIDE cases.

The prediction skill of IITM CFSv2 in predicting the considered MJO events is about 15 days, and the skill for LIDE and HIDE cases is around 14 and 15 days, respectively. i.e., all three sets of events show a similar range of prediction skill irrespective of the initial day error.

Fig 3 shows the growth of error with lead days for the LIDE and HIDE cases. This error growth is nothing but the difference of bivariate error in PC of $(n+1)^{th}$ and $n^{th}$ day forecasts. The error growth during the initial few days of the forecast is high for the LIDE cases compared to the HIDE cases. But after around a week, the error growth falls to the same range for both cases, highlighting that initial day error is not carried over to lead days more than ten days.

A notable difference is observed in the peaks of the probability density function curve of BVRMSE in PC for LIDE and HIDE cases during both the initial ten days and 10-20 days lead, as shown in Fig (4.a) and (4.b). During the initial ten lead days, the peak is different for both LIDE and HIDE cases (Fig 4.a). While considering the lead time of 10-20 days, the distribution seems to be broader for both sets of events (Fig 4.b). However, this difference in error distribution pattern does not seem to impact the probability density function curve of the error growth of bivariate PCs, which is similar for LIDE and HIDE, as can be seen from Fig (4.c) and (4.d). From Fig 4, it is evident that, although LIDE and HIDE cases represent the two extremes in the initial day error, their error growth patterns are the same. It infers that the error growth in IITM CFSv2 does not depend upon the initial day error for predicting the strong MJO events during MJJAS. Whatever the error in the initial day forecast, whether large or small, it does not influence the way error grows with lead days and has no effect on the forecast skill of the model.

**4.2       Model error growth pattern for LIDE and HIDE events with different initial phase**

The RMSE of PC amplitude of the LIDE and HIDE cases are calculated, and the events with different initial phases are identified. Further, the variation of RMSE of PC amplitude for these events with different initial phases up to a lead time of 32 days is given in Fig (5.a) and (5.b). A primary peak in error is seen for those events with the initial phase over the western Pacific (WP) (Phase 6) for the LIDE cases, which is quite extended for a lead time of 10-25 days. A secondary peak in error is also noticeable for events with the initial phase over parts of the Indian Ocean (IO) (Phase 2) and Maritime Continent (Phase 4). It shows that the events with initial phase lying over WP and IO contribute substantially towards the rapid growth of error with the lead time for the LIDE cases. For the RMSE of PC amplitude for the HIDE case shown in Fig (5.b), a primary peak is present for events with the initial phase over the IO (Phase 2 and 3). A secondary peak is also noted for events with an initial phase over the Maritime Continent. From the figure, it is clear that the Indian Ocean simulation of CFSv2 is the major contributor to error growth with lead days for the HIDE and partly in LIDE cases.

Even though the probability density function of error growth for LIDE and HIDE cases are found to be similar during 0-10, and 10-20 lead days, their density distribution of error shows slight differences. This variation can be attributed to the difference in phase distribution of error for the LIDE and HIDE cases (Fig 5). It makes the forecast of HIDE cases to fail even at the initial lead. For a better insight into the error distribution in LIDE and HIDE cases, we analyse the spatial distribution of convection and wind circulation anomalies between $15^0$N-$15^0$S. The OLR anomalies are regressed using the corresponding PC amplitudes, and the composite is taken for each pentad up to the fourth pentad. Along with the convection, dynamics associated with MJO propagation is analysed using the zonal wind profile (200hPa and 850hPa). A comparison of observational and model data of the same for the LIDE and HIDE is plotted in Fig 6 and 7 (discussed in next subsection).

### 4.3   Spatial distribution of the error growth in the OLR and circulation parameters for LIDE and HIDE events

The zonal wind anomaly at 850hPa level is shown in Fig 6 for the LIDE cases (Fig 6.a, 6.b, and 6.c) and HIDE cases (Fig 6.d, 6.e, and 6.f) along with the contours indicating OLR anomaly. In the spatial distribution of U850 wind from observation for the LIDE case (Fig 6.a) during lead time 1-5 days, the negative wind anomaly widespread over the Pacific Ocean indicates low-level easterly wind flow, and the positive wind anomaly over the western hemisphere and parts of African regions indicate a low-level westerly wind flow. The low level easterly to the right and westerly to the left indicates convection in the IO region, confirmed by the negative OLR anomaly persisting over IO (broken contour lines). The $2^{nd}$ pentad onwards propagation of low level westerly towards the Pacific Ocean can also be seen, indicating the eastward propagation of convective wind circulation. While considering the HIDE cases, we can infer from Fig (6.d) that the low-level westerlies to the right and strong easterlies to the left are present over the IO region, which clearly indicates the presence of an intense suppressed convection persisting over the region. This suppressed convection also exhibits eastward propagation during the subsequent pentads. The predicted low-level winds for the LIDE cases are underestimated (Fig 6.b and 6.c).However, for the HIDE cases , the circulation at the lower levels associated with suppressed convection over IO is very poorly captured by the model (Fig 6.e and 6.f).

The 200hpa circulation patterns are consistent with the 850hPa zonal wind flow (Fig 7); Upper-level easterlies over the western hemisphere and African regions (left) and strong westerlies over the Pacific Ocean (right) indicates upper-level divergence associated with the lower- level convergence and convection over IO region for the LIDE case (Fig 7.a).The reverse condition of upper level westerlies to the left (over African and Indian Ocean regions) and easterlies to the right (over the Pacific Ocean) indicates upper level convergence, which paves the way for subsidence and thereby suppressed convection to persist over the IO for the HIDE cases (Fig 7.d). Here also the model underestimates the magnitude of wind for the LIDE cases (Fig 7.b). From the difference plot (Fig 7.c), we can see that the peak error is found over the WP and then in the IO as was already inferred from Fig (5.a). For the HIDE cases (Fig 7.e), the westerlies over African and Indian Ocean regions are showing a dominant error, which is quite clear from the difference plot (Fig 7.f), accordant to Fig (5.b).

The OLR anomalies, together with the low level and upper level wind patterns, make it clear that the strong MJO events with convection over IO during the initial days are captured better than those with suppressed convection over IO. Despite a better forecast during the initial days for LIDE cases, the model fails to sustain this skill because of the inefficiency in capturing convection over IO and the circulation anomalies in both IO and WP. The poor simulation of suppressed convection over the IO makes the HIDE cases worsen even for the initial model forecast. For improving the skill of model it is necessary to give prime importance to the correction of model physics (rather than focusing ICs) that helps better simulation of wind and convection over Indian Ocean.

## 5. Discussions and Conclusions

The extended range forecasts of tropical intra-seasonal oscillations and MJO in a seamless framework have gained significant importance in the present scenario of the frequent occurrence of extreme events related to the large scale predictable component of the flow in the tropics. In a seamless forecast approach, although theoretically it is expected that improvement in skill of extended range forecast is related with the availability of improved initial condition, it is not clear if the MJO or the extremes associated with MJO could be forecasted in a better way in the operational extended range scale if the initial MJO phase has low error. In the present work, we have evaluated this hypothesis if the growth of error in the extended range scale is related to the initial forecast error using the extended range MJO forecast from the IITM CFSv2 model during MJJAS. Apart from this we also attempted to identify the physical attribute inducing these errors separating the strong MJO events into two set of events (Section 3.3) with the least initial day error (LIDE) and the highest initial day error (HIDE). It is clearly noted (Section (4.1)) that the prediction skill of IITM CFSv2 model for LIDE and HIDE cases is around 15 days. Further, it was seen that irrespective of the error nature in the initial day the error growth in tracking MJO is comparable after a lead of 7-10 days for both LIDE and HIDE cases (Fig 3). It is explicit in the probability density function (pdf) curve of error and error growth (Fig 4). The slight variation in the pdf of error was not affecting the pdf of error growth. While the LIDE cases show peak error for events with initial phase in the WP and the IO, the HIDE cases, on the other hand, show peak error for events with initial phase in the IO region (Fig 5) The spatial distribution of OLR and circulation anomalies for LIDE and HIDE cases (Fig 6 and 7) revealed that the capability of the model in capturing convection over IO is less and this has a major role in the error growth of model forecast. Though LIDE cases initially performs well, they fail during longer lead because the convection and the circulation features over WP and IO are inappropriately captured. While for HIDE cases the suppressed convection over IO is very poorly captured from the initial leads.

There is a widely accepted belief that improving the initial conditions (i.e., minimizing the initial error) can improve the performance of a climate model. Although this is valid for the initial few days of the forecast, the effect of initial error in the forecasts of lead time greater than ten days is much limited while predicting MJO using IITM CFSv2 model. Whether the initial condition used for the forecast has less or more error, the model forecast of MJO after ten days is having a similar error. The major factor affecting the prediction skill in the extended range is the constraints in model physics. Whether it is the LIDE or HIDE case, the incorrect model forecast primarily over the IO and WP region is what that reduces the prediction skill of the IITM CFSv2 model. This model performance over the Indian Ocean has to be improved for better capturing of MJO irrespective of the error in the initial day of forecast.

It is to be mentioned, however, that in this analysis the role of ICs are not ignored. But the information of error in PC is not carried beyond a lead time of ten days in the IITM-CFSv2 extended range forecast. The purpose of extended range prediction is to minimize the error in model forecast beyond 10 days. For that more emphasis is to be given in sorting out the biases in model dynamics and physics rather than focusing only on the ICs. Here for the IITM CFSv2 model forecasts during MJJAS the model physics has to be revised for better capturing of convection over IO in addition to improving the ICs for the overall enhanced performance of the model.

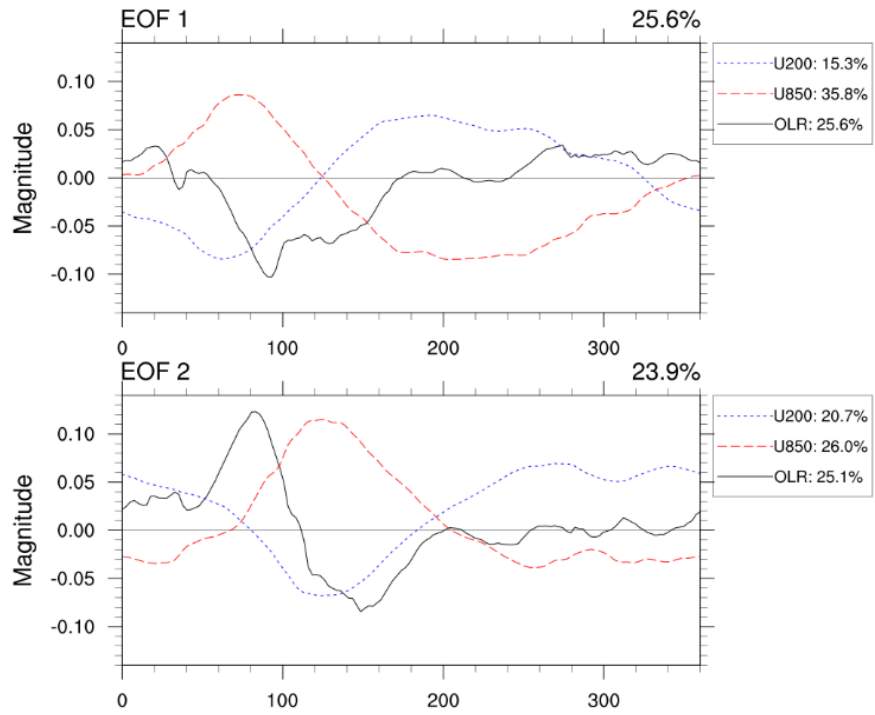

*Figure 1*: Spatial structure of leading pair of EOFs obtained using observational OLR, u850 and u200 fields and the explained variance &percentage of variance associated with each variable

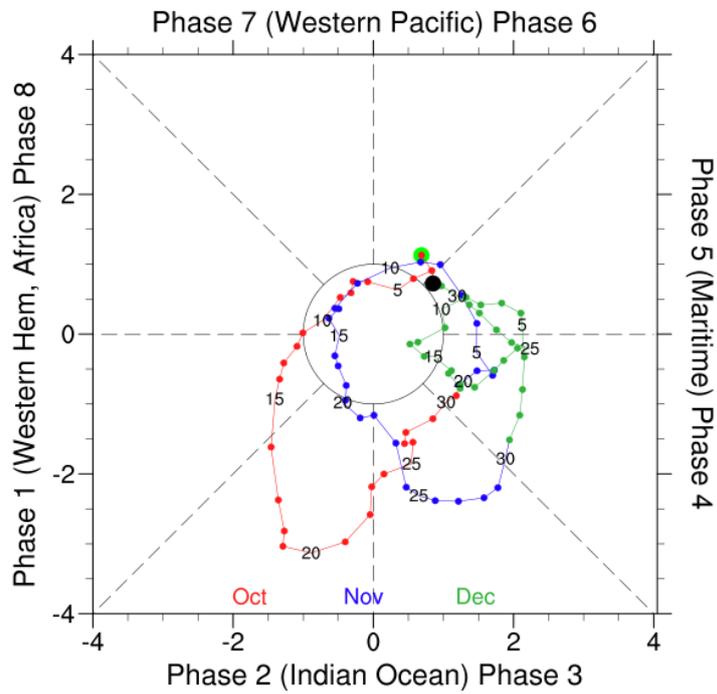

***Figure 2*** *: Phase plot of the DYNAMO period (Oct-Nov 2011) of MJO. The X and Y axes are RMM1 and RMM2 respectively. The MJO signals lying inside the unit circle are considered to be weak. The green dot shows the starting date with each month given a different colour and the end date is indicated by the black dot.*

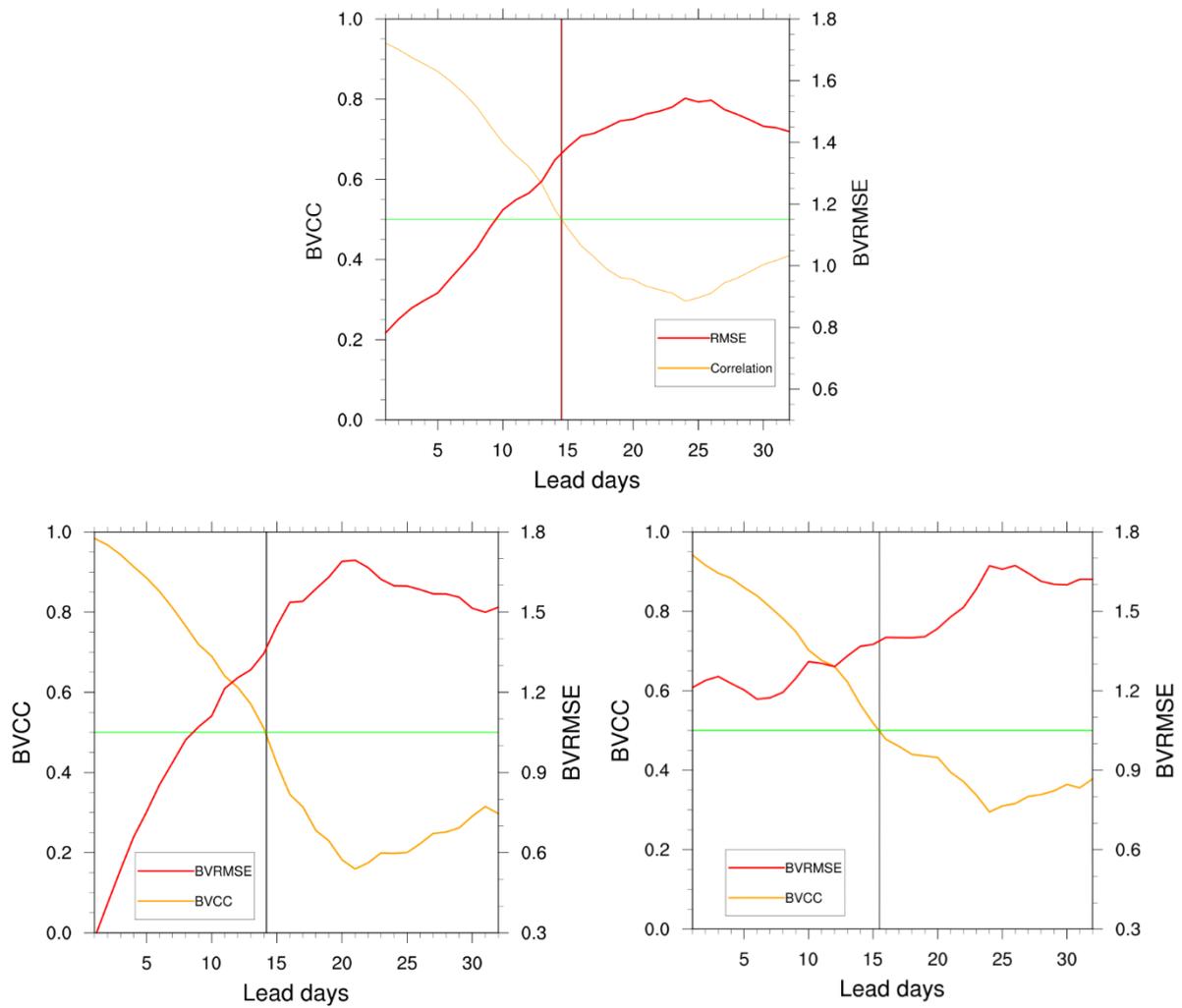

**Figure 3**: BVRMSE and BVCC of MJO Events as predicted using IITM CFSv2 model from 2003-2018 MJJAS. The green line represents the correlation value of 0.5 which when intersects the BVCC (orange) line gives the prediction skill of model. (a) BVRMSE and BVCC of all strong MJO events (b) BVRMSE and BVCC of LIDE cases (c) BVRMSE and BVCC of HIDE cases

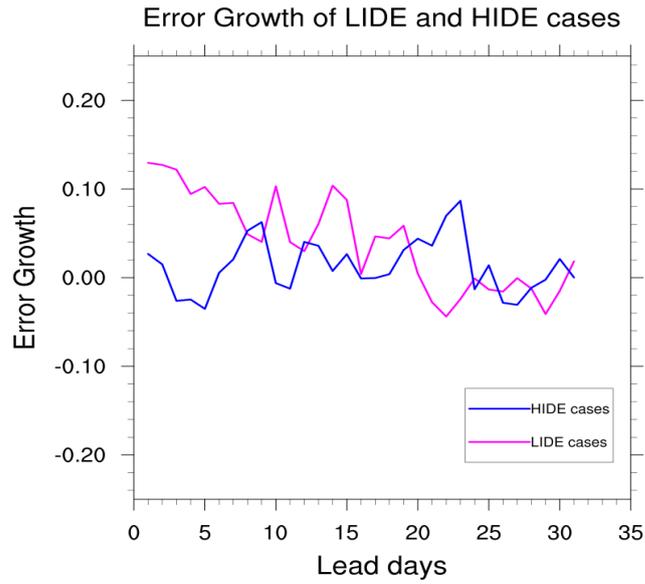

*Figure 4*: Growth of BVRMSE of the LIDE and HIDE cases with lead days. Blue line shows the error growth of HIDE cases and magenta line shows that of LIDE cases.

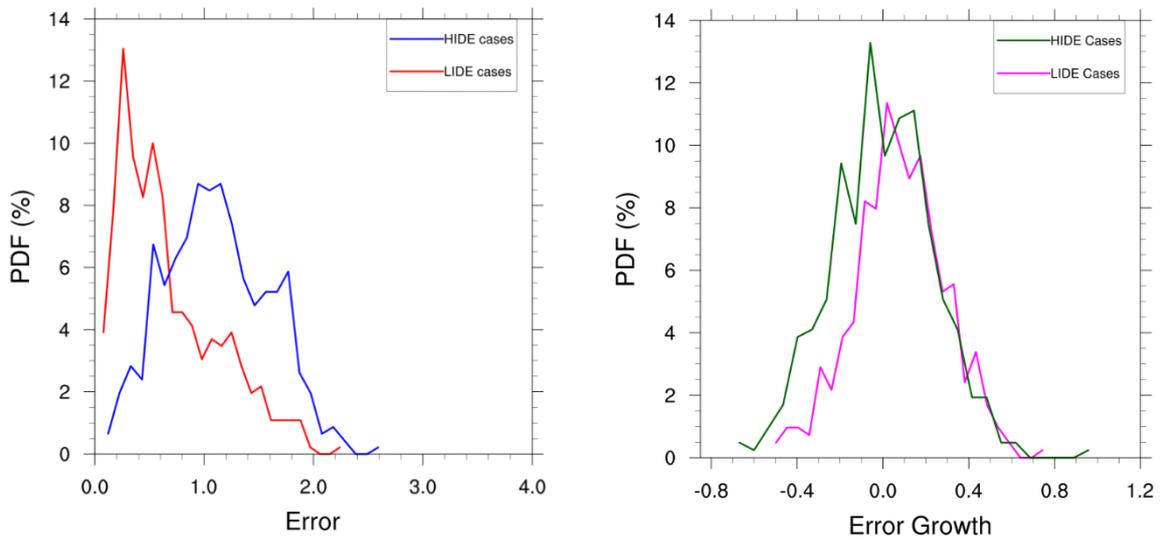

*Figure 5*: Probability density function curve of (a) BVRMSE and (b) growth of BVRMSE for LIDE and HIDE cases for 1-10 days lead.

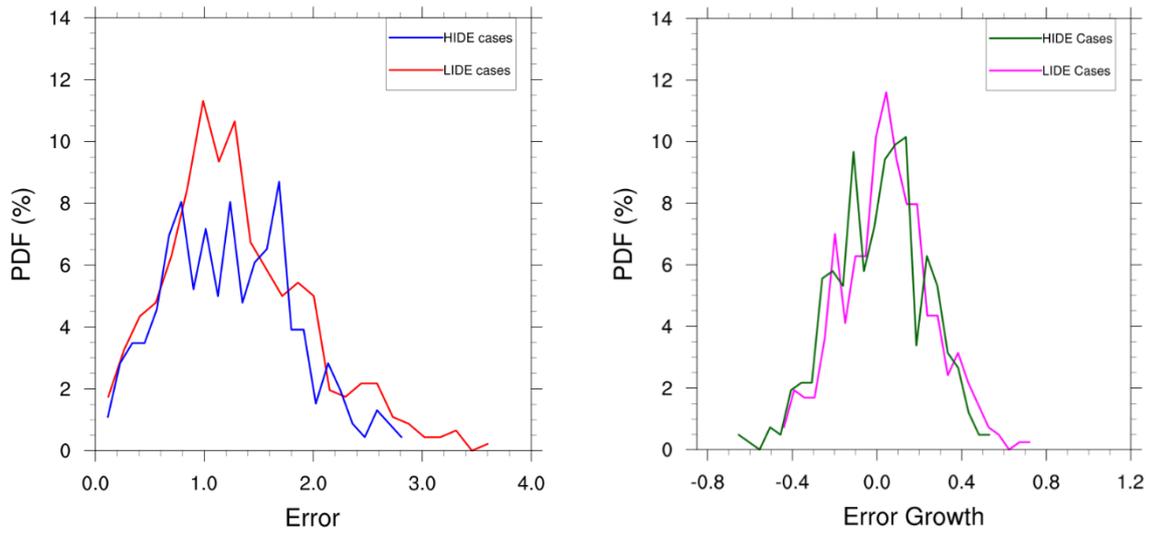

*Figure 6*: *Probability density function curve of (a) BVRMSE and (b) growth of BVRMSE for LIDE and HIDE cases for 10-20 days lead.*

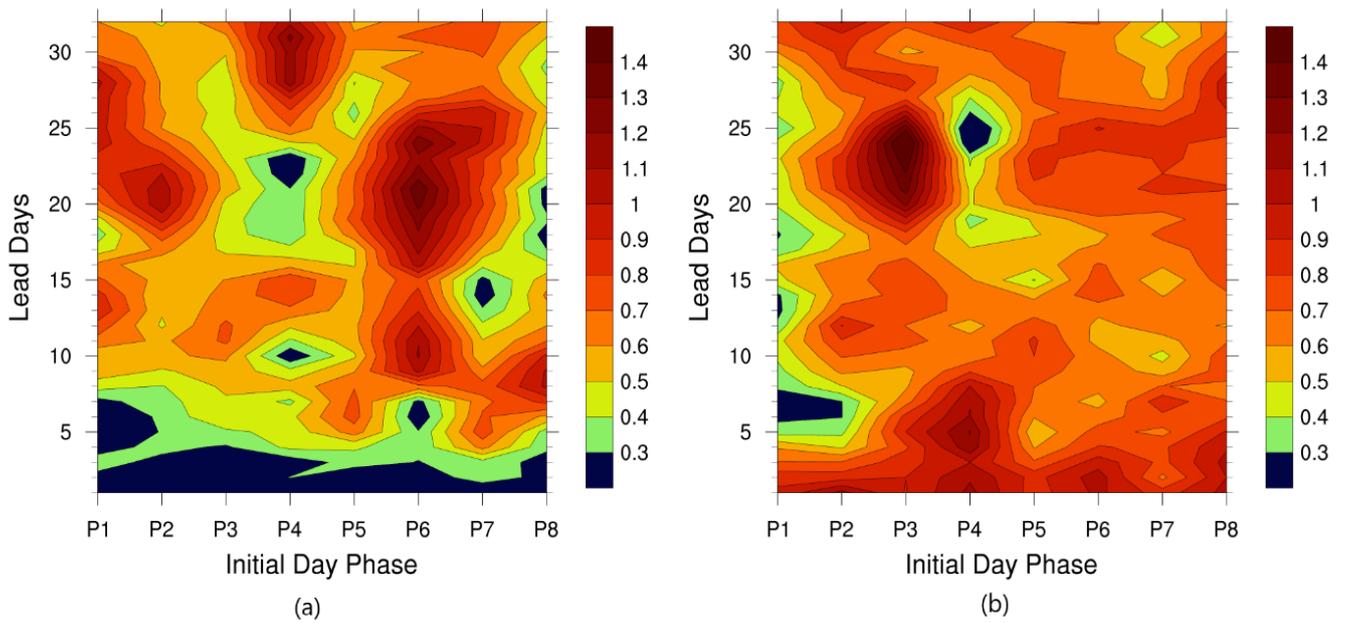

*Figure 7*: *Root mean square error of bivariate PC amplitude as a function of initial day phase and lead time. (a) for LIDE cases (b) for HIDE cases*

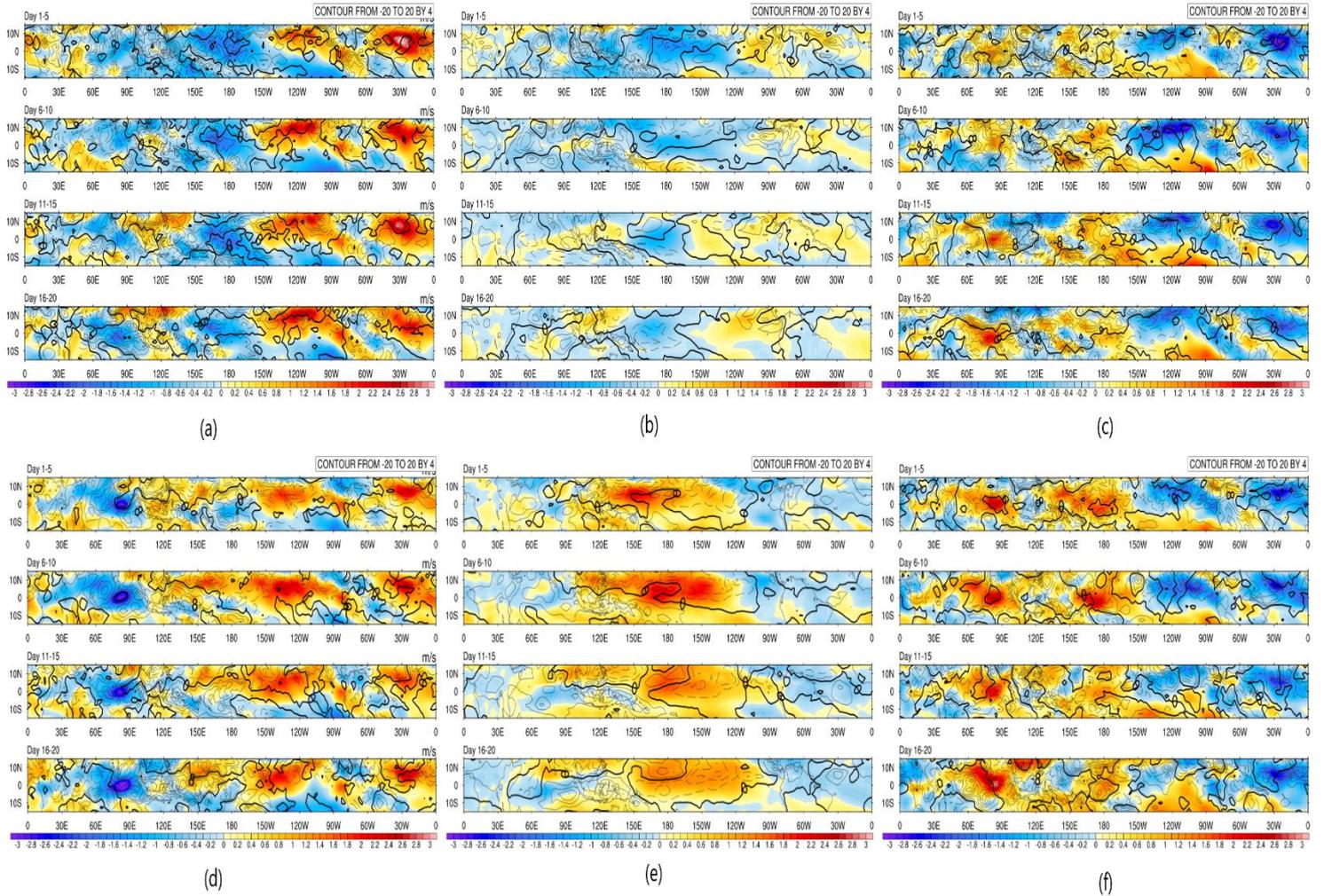

***Figure 8***: *Spatial distribution of zonal winds at 850 hPa level for the LIDE from the observational data, model data and the difference between model and observation (Model-Observation) respectively are shown in the first row. The same for HIDE cases are shown in the second row. The contour lines represent the OLR anomalies and positive anomaly is shown by solid lines and negative anomaly is shown using broken lines.*

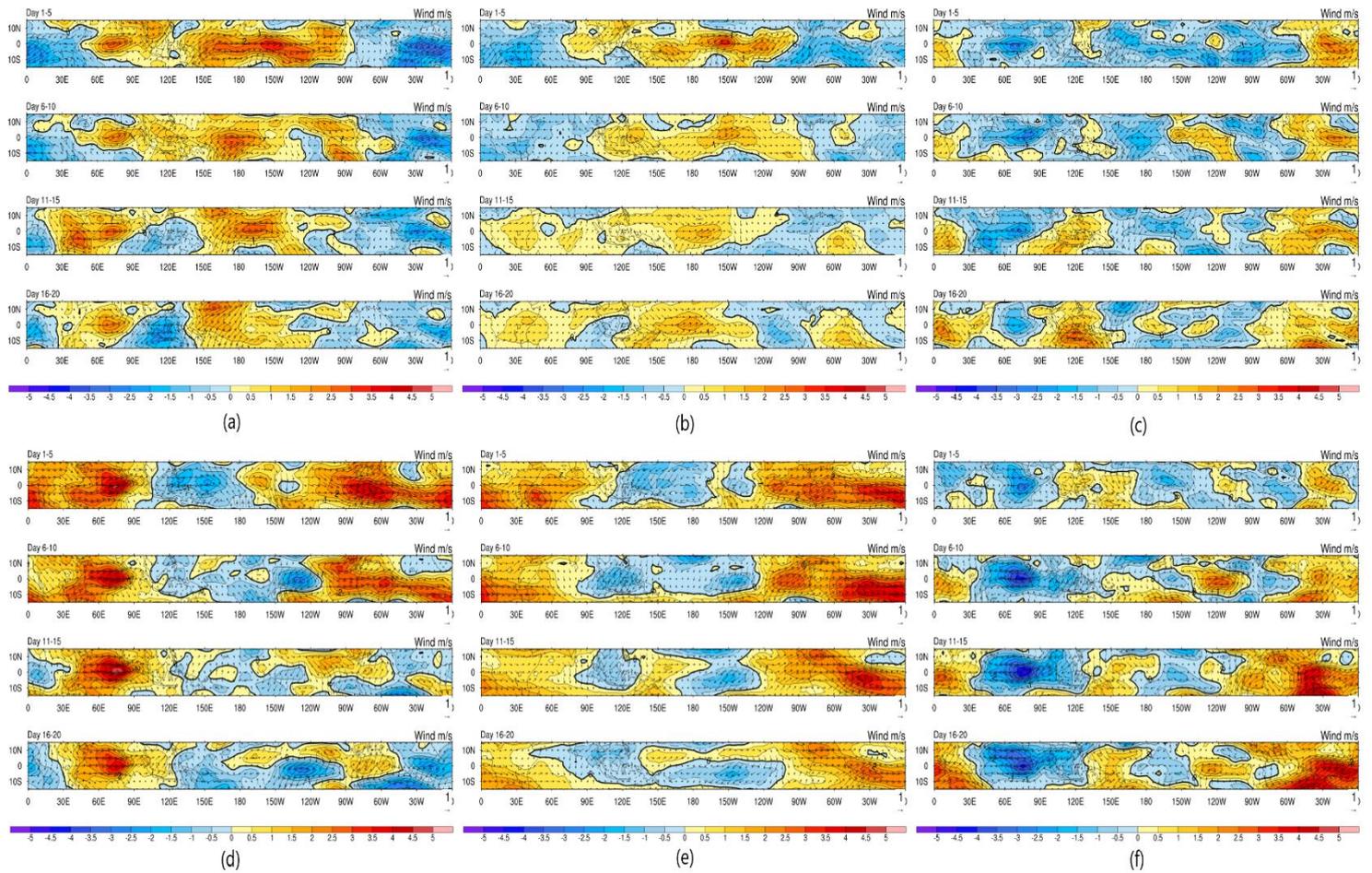

*Figure 9*: *Spatial distribution of zonal winds at 200 hPa level for the LIDE from the observational data, model data and the difference between model and observation (Model-Observation) respectively are shown in the first row. The same for HIDE cases are shown in the second row. The vector wind at 200 hPa level is shown by the black arrows.*